\documentclass[12pt,twoside]{article}
\usepackage{graphicx}
\usepackage{amssymb}
\usepackage{amsmath}
\begin{document}
	
\def\bd{\begin{description}}
\def\be{\begin{enumerate}}
\def\ben{\begin{equation}}
\def\benn{\begin{equation*}}
\def\een{\end{equation}}
\def\eenn{\end{equation*}}
\def\benr{\begin{eqnarray}}
\def\eenr{\end{eqnarray}}
\def\benrr{\begin{eqnarray*}}
\def\eenrr{\end{eqnarray*}}
\def\ed{\end{description}}
\def\ee{\end{enumerate}}
\def\al{\alpha}
\def\b{\beta}
\def\bR{\bar\R}
\def\bc{\begin{center}}
\def\ec{\end{center}}
\def\d{\dot}
\def\D{\Delta}
\def\del{\delta}
\def\ep{\epsilon}
\def\g{\gamma}
\def\G{\Gamma}
\def\h{\hat}
\def\iny{\infty}
\def\La{\Longrightarrow}
\def\la{\lambda}
\def\m{\mu}
\def\n{\nu}
\def\noi{\noindent}
\def\Om{\Omega}
\def\om{\omega}
\def\p{\psi}
\def\pr{\prime}
\def\r{\ref}
\def\R{{\bf R}}
\def\ra{\rightarrow}
\def\s{\sum_{i=1}^n}
\def\si{\sigma}
\def\Si{\Sigma}
\def\t{\tau}
\def\th{\theta}
\def\Th{\Theta}
\def\vep{\varepsilon}
\def\vp{\varphi}
\def\pa{\partial}
\def\un{\underline}
\def\ov{\overline}
\def\fr{\frac}
\def\sq{\sqrt}
\def\WW{\begin{stack}{\circle \\ W}\end{stack}}
\def\ww{\begin{stack}{\circle \\ w}\end{stack}}
\def\st{\stackrel}
\def\Ra{\Rightarrow}
\def\R{{\mathbb R}}
\def\bi{\begin{itemize}}
\def\ei{\end{itemize}}
\def\i{\item}
\def\bt{\begin{tabular}}
\def\et{\end{tabular}}
\def\lf{\leftarrow}
\def\nn{\nonumber}
\def\va{\vartheta}
\def\wh{\widehat}
\def\vs{\vspace}
\def\Lam{\Lambda}
\def\sm{\setminus}
\def\ba{\begin{array}}
\def\ea{\end{array}}
\def\bd{\begin{description}}
\def\ed{\end{description}}
\def\lan{\langle}
\def\ran{\rangle}

\title{ Squeezing and Entanglement of two-modes  Quantum $\mathrm{X}$ Waves}

\author{Ali Saif M. Hassan$^{1,}$\footnote{Electronic address: alisaif73@gmail.com},  Waleed S. A. Hasan$^{1,2}$\footnote{Electronic address: waleedj166@gmail.com}. and  M. A.  Shukri$^{2,}$\footnote{Electronic address: mshukri2006@gmail.com}\\$^1$ Department of Physics, University of Amran, Amran, Yemen.\\ $^2$ Department of Physics, University of Sana'a, Sana'a, Yemen.}
\maketitle

\begin{abstract}
	quantum theory of generalized $\mathrm{X}$ waves with orbital angular momentum in dispersive media, and the interaction of quantized $\mathrm{X}$ waves in quadratic  nonlinear media were studied in (J. opt,20,065201(2018)). We present a kind of phase matching, which is called velocity phase matching, and this  phase matching can be used for determining the length of the nonlinear crystal or the interaction time in the experiment setup, to produce $\mathrm{X}$ waves with particular velocity $v$. Moreover, we introduce more analysis for the dependence of squeezing of $\mathrm{X}$ waves on its spectral order, and for spectral orders $j>0$, we predict  the existence of a characteristic axicon aperture for maximal squeezing. Then we find the quantum squeezed state of down-converted state generated by the $\chi^{2}$-nonlinear process. Finally, we detect their entanglement  using a criterion of separability.
\end{abstract}
\section{Introduction} Squeezed states, and two-mode-squeezed states, are foundational for describing the basis nature of quantum mechanical phenomena. their quantum entanglement plays a central role in quantum physics, experiments of quantum optics, and quantum information science [1-6]. Nonlinear quantum optics, in particular, parametric down-conversion {\bf PDC}, gives us ways to study issues in the foundations of quantum mechanics that are not readily available using other technologies \cite{Reid,AD}. The quantum communication encode information into the polarization degrees of freedom of photons \cite{D} or the orbital angular momentum {\bf OAM} \cite{E,F,G,H,I,J,K}.The important problem related to multi-modes quantum communications is the diffraction and dispersion of the electromagnetic waves \cite{F}. Electromagnetic waves packet is usually subjected to diffraction and dispersion. Diffraction create a broadening in space during propagation of the wave and dispersion create a broadening in the time during the propagation. Ultimately, these effects are connected with the bounded nature of the wave spectrum and, therefore, to its finite energy content \cite{L}. A great effort has been done in studying the effect of atmosphere turbulence in free-space communication \cite{G,M,N,O,P}. Maxwell's equations admit diffraction-free and dispersion-free solutions, the so-called localized waves \cite{Q}. In particular, such solutions in the monochromatic domain are the Bessel beams \cite{R}, and in the pulsed domain, the most renowned localized waves are the $\mathrm{X}$ waves, first introduced in acoustics in 1992 by Lu and GreenLeef \cite{S,T}. Despite the great work in literature concerning $\mathrm{X}$ waves, the investigations of their quantum properties are very few \cite{U,V}. Generalization of the traditional $\mathrm{X}$ waves to the case of {\bf OAM}-carrying $\mathrm{X}$ wave and the coupling between angular momentum and the temporal degrees of freedom of ultrashort pulses have been investigated \cite{Y}. Very recently, quantum and squeezing of $\mathrm{X}$ waves  with {\bf OAM}  in nonlinear dispersive media have been proposed  \cite{orn17,orn18}, and they can open a new direction for free-space quantum communication and different areas of physics.
 In this work, we state the propagation of a scalar electromagnetic field in a linear dispersive medium. We present a kind of phase matching called velocity phase matching and introduce a relation between the interaction time and the velocity of $\mathrm{X}$ wave as well as a relation between  the length of the $\chi^{(2)}$-nonlinear crystal and the velocity of $\mathrm{X}$ wave generated by the spontaneous parametric down-conversion {\bf SPDC} process  in Sec. 2.  We study the  {\bf SPDC} process in a quadratic medium, in particular the dependence of squeezing of the down-converted state generated by the $\chi^{(2)}$-nonlinear crystal on the spectral order of quantum  $\mathrm{X}$ waves in Sec. 3. We introduce the quantum squeezed form of the state generated by the {\bf SPDC} process and its entanglement in Sec. 4. finally, the results are summarized in Sec. 5.

 \section{Propagation of a generalized $\mathrm{X}$ wave in  dispersive media and its quantization }
  We consider the propagation of a scalar electromagnetic field in a linear dispersive medium with refractive index $n=n(\omega)$. Applying the paraxial and slowly varying envelope approximation to the envelope function  $A({\bf r},t)$ of an electric field \begin{equation} E({\bf r},t) = \sqrt{\frac{2}{\epsilon_{0}n^{2}}}A({\bf r},t) e^{i(k z -w t)},\end{equation}  where $A({\bf r},t)$ varies slowly with $z$, and $k=n(\omega)\omega/c,$ which are the propagation coordinate and the wave number respectively.\\ The field envelope $A({\bf r},t)$ satisfies the following equation \cite{orn18}, \begin{equation}\label{eq:eps} i\frac{\partial A}{\partial t} +i\omega'\frac{\partial A}{\partial z} - \frac{\omega''}{2} \frac{\partial^{2} A}{\partial z^{2}} +\frac{\omega'}{2k}\Delta^{2}_{\bot} A =0, \end{equation}   where $\omega'=c^2 dk/d\omega$ and  $\omega''=c^2 d^2k/d\omega^2$ are the first and the second order dispersion respectively.\\ The solution of Eq.(\ref{eq:eps}) can be written as  a superposition of the generalized  $\mathrm{X}$ waves as follows \cite{orn18}:
\begin{equation}\label{a} A({\bf r},t) = \sum_{m,p}\int dv\; C_{mp}(v)\; e^{-i\frac{v^{2}}{2\omega''}t}\;  \psi^{(v)}_{m,p}(R,\zeta), \end{equation}  where $\psi^{(v)}_{m,p}(R,\zeta)$ is the generalized $\mathrm{X}$ wave of {\bf OAM} number $m,$ spectral order $p$ and velocity $v$ defined as, \begin{equation}
\psi_{m,p}^{(v)}(R,\zeta)=\int_{0}^{\infty} d\alpha f_{p}(\alpha) J_{m}(\sqrt{\omega''k/\omega'}\alpha R)\; e^{i(\alpha- v/\omega'')\zeta} \; e^{im\theta},
\end{equation}  with $R=\sqrt{x^{2}+ y^{2}}$, $\zeta=z- (\omega'+v) t$  which is the co-moving  reference frame associated to the $\mathrm{X}$ waves, and
$f_{p}(\alpha)=\sqrt{k /\pi^{2}\omega'(1+p)} (\Delta \alpha) L_{p}^{(1)}(2\alpha\Delta) e^{-\alpha\Delta}$ which is the spectrum function, where $L_{p}^{(1)} (x)$ is the generalized  Laguerre  polynomials  of the first kind of order $p$ \cite{Z}, and $\Delta$ is the  reference length related to the spatial extension of the spectrum \cite{orn18}. The quantized field of Eq.(\ref{a}) can de written as  \cite{orn18},
\begin{equation}
\hat{A}({\bf r},t)=\sum_{m,p}^{\infty}\int dv\; e^{ \frac{-i}{\hbar}(\frac{M v^{2}}{2}) t} \sqrt{\hbar\omega_{m,p}(v)}\;\psi^{(v)}_{m,p}(R,\zeta)\;\hat{a}_{m,p}(v) + h.c.,
\end{equation} where $h.c.$ denotes to the Hermitian conjugate and $M= \hbar/\omega''$ is the mass of $\mathrm{X}$ wave. The Hamiltonian operator can be written as follows \cite{orn18}: \begin{equation}\label{aa}
\hat{H}=\sum_{m,p}\int dv\;{\hbar}\omega_{m,p}(v)\Big(\hat{a}^\dagger_{m.p}(v)\;\hat{a}_{m,p}(v)+\frac{1}{2}\Big),
\end{equation} with the following usual bosonic commutation relations of creation and annihilation operators of the field, $$[\hat{a}_{m,p}(v),\hat{a}^\dagger_{n,q}(u)]=\delta_{m,n}\;\delta_{p,q}\;\delta(u-v),$$
\begin{equation} \label{anh}[\hat{a}_{m,p}(v),\hat{a}_{n,q}(u)]=0=[\hat{a}^\dagger_{m,p}(v),\hat{a}^\dagger_{n,q}(u)].\end{equation}

Now we turn our attention to the quadratic nonlinear process involving  $\mathrm{X}$ waves  particularly to the spontaneous parametric down conversion {\bf SPDC} \cite{AD}.
The interaction Hamiltonian of this process, with a practical case $\rho=\sqrt{k_{1}\omega'_{2}/k_{2}\omega'_{1}} \simeq 1,$ can be written in two different forms  \cite{orn18}:\\
 The first form, is for the time  dependent interaction Hamiltonian which can be written as follows \cite{orn18}:
\ben\label{hmi}\hat{H}_{I}(t)=\hbar\sum_{m,p,q}\int du\; dv\; \chi_{m,p,q}(u+v)\sqrt{\omega_{m,p}(u)\omega_{-m,p}(v)}\; e^{iF(u,v)t} \;\hat{a}_{m,p}^\dagger(u)\;\hat{b}_{-m,q}^\dagger(v)+h.c.,\een
with\ben F(u,v)= \frac{1}{2\omega''}[\;2uv + (v-u)(\omega'_{1}-\omega'_{2})],\een and the interaction function is  \ben \label{int}\chi_{m,p,q}(u+v)=\frac{(-1)^m\; \chi\; \Delta^{2}}{\omega''^{2} \sqrt{(1+p)(1+q)}} (u+v)\;  L_{p}^{(1)}(\frac{(u+v)\Delta}{2\omega''})\;L_{q}^{(1)}(\frac{(u+v) \Delta}{2\omega''})\;e^{\frac{-(u+v)\Delta}{2\omega''}}\; \theta({u+v}),\een where $\theta({u+v})$, is the Heaviside step function  \cite{Z}.
The second form, is for the finite length interaction Hamiltonian which can be given by \cite{orn18}  \begin{equation}\hat{H}_{I}(z)=\int dt\; [\;\chi\; \langle A_{1}|A^{\ast}_{2}\rangle + \chi^{\ast}\langle A_{2}|A^{\ast}_{1}\rangle]. \end{equation}  The final form of $\hat{H}_{I}(z)$ can be written as follows:\ben\label{b}\hat{H}_{I}(z)=\hbar\sum_{m,p,q}\int du\;dv\; \Xi_{m,p,q}(u,v)\;\sqrt{\omega_{m,p}(u) \;\omega_{-m,q}(v)}\; e^{i\Lambda(u,v)z}\;   \hat{a}_{m,p}^\dagger(u)\;\hat{b}_{-m,q}^\dagger(v)+h.c.,\een
where $\Lambda(u,v)=2uv/\omega''(u+v),$  and  $\Xi_{m,p,q}(u,v),$ which is the modified vertex function, can be given by\cite{orn18},
\ben\label{vf} \Xi_{m,p,q}(u,v)=\frac{(-1)^{m}\; 2\;\chi\; \Delta^{2} }{\omega''^{2} \sqrt{(1+p)(1+q)}}\frac{u^{2}+v^{2}}{u+v}\; L_{p}^{(1)}(\frac{(u^{2}+v^{2})\Delta}{(u+v)\omega''})\; L_{q}^{(1)}(\frac{(u^{2}+v^{2})\Delta}{(u+v)\omega''})\; e^{-\frac{(u^{2}+v^{2})\Delta}{(u+v)\omega''}}.\een
To calculate the state of the system after the interaction, we apply Schwing-Dyson expansion of the propagator $\exp[-i\hat{H}_{I}(t)/\hbar]$, and consider only the first term of the expansion.
For the time dependent form of the interaction Hamiltonian stated in Eq.(\ref{hmi}), the final state after the $\chi^{(2)}$-nonlinear interaction can be expressed as follows \cite{orn18}:
\ben  |\psi^{(1)}(t)\rangle=\frac{-i}{\hbar}\int_{0}^{t}d\tau \;\hat{H}_{I}(\tau)|0\rangle. \een  Substituting Eq.(\ref{hmi}) into the above equation and after some calculations, the final state can be written in the following form:
\ben \label{stt1} |\psi^{(1)}(t)\rangle=\sum_{m,p,q} \int du \; dv \; \mathcal{G}_{m,p,q}(u,v)\; \mathcal{F}(u,v,t) \;|m,p,u;-m,q,v\rangle,\een
where $|m,p,u;-m,q,v\rangle\equiv \hat{a}^{\dagger}_{m,p}(u)\;\hat{b}^{\dagger}_{-m,q}(v)|0,0\rangle,$ the amplitude $\mathcal{G}_{m,p,q}(u,v)$ is defined as
\ben \mathcal{G}_{m,p,q}(u,v)= -i\sqrt{\omega_{m,p}(u)\;\omega_{-m,q}(v)}\;\frac{\chi_{m,p,q}(u+v)}{F(u,v)},\een
and $\mathcal{F}(u,v,t)=[e^{iF(u,v)t}-1].$
The above state represents the superposition of two particles corresponding to the modes $\omega_{1}$ and $\omega_{2}$  (generated by the nonlinear interaction) traveling with velocities $u$ and  $v$ respectively. For the second form of the interaction Hamiltonian stated in Eq.(\ref{b}), the final state of the system can be written as
\ben |\psi^{(1)}(L)\rangle=\frac{-i}{\hbar}\int_{0}^{L}dz \;\hat{H}_{I}(z)|0\rangle.\een
 Substituting  Eq.(\ref{b}) into the above equation, the state of the field at the output faced of the crystal can be written as follows:
\ben |\psi^{(1)}(L)\rangle=\sum_{m,p,q}\int du \; dv \;\mathcal{L}_{m,p,q}(u,v)\;  \mathcal{F}(u,v,L)|m,p,u;-m,q,v\rangle,\een
where $\mathcal{F}(u,v,L)=[e^{i\Lambda(u,v) L}-1],$ and \ben \mathcal{L}_{m,p,q}(u,v)=\frac{-i}{2}\sqrt{\omega_{m,p}(u)\omega_{-m,q}(v)}\;[\frac{\omega''(u+v)}{uv}] \; \Xi_{m,p,q}(u,v). \een
The transition probability for the field to be in such a state after the time dependent interaction with the $\chi^{(2)}$-nonlinear crystal is proportional with $|\mathcal{F}(u,v,t)|^{2}$ as follows:
\ben P(t)\propto \vert \mathcal{F}(u,v,t)\vert^{2}, \een
similarly, the transition probability for the finite length interaction of the $\chi^{(2)}$-nonlinear crystal is proportional with $|\mathcal{F}(u,v,L)|^{2}$ as follows \cite{orn18}:
\ben P(L)\propto \vert \mathcal{F}(u,v,L)\vert^{2}.\een
Explicitly, $\vert \mathcal{F}(u,v,t)\vert^{2}$  can be expressed as follows:
\ben \label{ff} \vert \mathcal{F}(u,v,t)\vert^{2}=2[1-\cos (F(u,v)t)].\een
From  the above equation we notice that the transition probability reaches its maximum, when $\vert \mathcal{F}(u,v,t)\vert^{2}$
reaches its maximum, and that occurs  when the condition $ F(u,v)t =(2m+1)\pi$, with $m=0,1,2,3,\dots$, is achieved.
The above condition is considered as a new phase matching condition, called  velocity matching condition  over the velocities of the two-modes $\mathrm{X}$ waves generated from {\bf SPDC} process. This condition requires that \ben u=\frac{2k'_{m}+v \;\Delta \omega'}{2v+\Delta\omega'},\een where $\Delta \omega'=\omega'_{1}-\omega'_{2},$\; and $k'_{m}=(2m+1)\pi\omega''/t$. For the case, when $\Delta \omega'=0$, we get $uv= k'_{m} =(2m+1)\pi\omega''/t,$ and the interaction time can be given by $t=(2m+1)\pi\omega''/uv$.
When $u=v$, regardless of that  $\Delta \omega'=0,$  we get
$v^{2} = k'_{m}=(2m+1)\pi \omega''/t.$\\
Thus, the interaction time can be written in terms of the velocity of  $\mathrm{X}$ waves modes as follows:
\ben\label{ti}  t=\frac{(2m+1)\pi \omega''}{v^{2}}.\een
For the finite length of the nonlinear crystal, the velocity matching condition is given by \cite{orn18},
\ben u=\frac{k_{n}\;v}{v-k_{n}},\een
where
$k_{n}=(n+\frac{1}{2})\pi\omega''/L$, with $n= 0,1,2,\cdots.$ Therefore the velocity matching condition fixes either the length of the crystal $L$ or the relative velocity of the two modes $\mathrm{X}$ wave involved in the {\bf SPDC} process as, $uv/(u+v)= k_{n} = (n+1/2)\pi \omega''/L$, and the length of the crystal $L$ can be given by, \ben L=\frac{(n+\frac{1}{2})\pi \omega''\;(u+v)}{uv}.\een
For $u=v$, the above equation becomes
\ben\label{li}  L=\frac{(2n+1)\pi \omega''}{v}.\een
From Eqs.(\ref{ti}) and (\ref{li}), we note that the ordinary relation between the length and the time is achieved ($L=vt$). This result can be used for determining the length of the nonlinear crystal in the experimental setup, to produce $\mathrm{X}$ waves with velocity $v$. This is the first result of our work. \\
\section{Squeezing of quantum $\mathrm{X}$ waves}
Now we can study the general case of squeezing effect in the {\bf SPDC} process in particular, the dependence of squeezing of the down-converted state, generated by $\chi^{(2)}$-nonlinear media, on the spectral orders of quantum $\mathrm{X}$ waves, for the case, when the velocities of the two-modes $\mathrm{X}$ waves are equal  $u=v$, and their spectral orders are equal as well $p=q=j.$ The interaction time  Hamiltonian in Eq.(\ref{hmi}) becomes
\ben\label{hmti}\hat{H}_{I}(t)=\hbar\sum_{mj}\int dv\; \chi_{mj}(2v)\; \omega(v)\; \hat{a}_{mj}^\dagger(v,t)\;\hat{b}_{-mj}^\dagger(v,t)+h.c.,\een where $ \hat{a}_{mj}^\dagger(v,t)=e^{iF(v)t/2} \;\hat{a}_{mj}^\dagger(v),$ and $\hat{b}_{-mj}^\dagger(v,t)=e^{iF(v)t/2} \;\hat{b}_{-mj}^\dagger(v)$.
Also the interaction function in Eq.(\ref{int}) becomes
\ben\label{int2}\chi_{mj}(2v)=\frac{2(-1)^{m}\;\chi \Delta^{2} v}{\omega''^{2}(1+j)}\; L_{j}^{(1)}(\frac{v \Delta}{\omega''})\; L_{j}^{(1)}(\frac{v \Delta}{\omega''})\; e^{\frac{-v\Delta}{\omega''}}.\een
And the modified vertex function in Eq.(13) can be rewritten as follows:
\ben\label{vfm}\Xi_{mj}(v)=\frac{2(-1)^{m}\;\chi \Delta^{2}\;v}{\omega''^{2}(1+j)}\; L_{j}^{(1)}(\frac{v \Delta}{\omega''})\; L_{j}^{(1)}(\frac{v \Delta}{\omega''})\; e^{\frac{-v\Delta}{\omega''}}.\een
From Eqs.(\ref{int2}) and (\ref{vfm}) we notice that the interaction function and the modified vertex function have the same form.
In the interaction picture we consider the time evolution controlled by $\hat{H}_{I}(t)$. Thus, the equations of motion for $\hat{a}_{mj}(v,t)$ and $\hat{b}_{-mj}(v,t)$  are:
\ben \frac{d\hat{a}_{mj}(v,t)}{dt}=\frac{1}{i\hbar}[\hat{a}_{mj}(v,t),\hat{H}_{I}(t)],\een
\ben \frac{d\hat{b}_{-mj}(v,t)}{dt}=\frac{1}{i\hbar}\;[\hat{b}_{-mj}(v,t),\hat{H}_{I}(t)].\een
Substituting Eq.(\ref{hmti}) into the above equations, we obtain
\ben \frac{d\hat{a}_{mj}(v,t)}{dt}=-2i \;\omega(v)\;\chi_{mj}(2v)\;\hat{b}_{-mj}^{\dagger}(v,t),\een
\ben \frac{d\hat{b}_{-mj}(v,t)}{dt}=-2i \;\omega(v)\;\chi_{mj}(2v)\;\hat{a}_{mj}^{\dagger}(v,t).\een
The general solutions of these equations are \cite{mand},
\ben \hat{a}_{mj}(v,t)=\cosh (\beta_{mj}(v)\; t)\; \hat{a}_{mj}(v)+\sinh (\beta_{mj}(v)\; t)\; \frac{\alpha_{mj}}{\beta_{mj}(v)}\;\hat{b}_{-mj}^{\dagger}(v),\een
 \ben \hat{b}_{-mj}(v,t)= \cosh (\beta_{mj}(v)t) \hat{b}_{-mj}(v)+\sinh (\beta_{mj}(v)\; t)\; \frac{\alpha_{mj}}{\beta_{mj}(v)}\;\hat{a}_{mj}^{\dagger}(v).\een
Here, we consider the expression $\alpha_{mj}=-2i\;\omega(v)\;\chi_{mj}(2v)$, with $\omega(v)= v^2/2\omega''.$ To find the squeezing parameter $\xi_{mj}(v)=\beta_{mj}(v) t$, we use the relation $\beta_{mj}(v)=\sqrt{\alpha_{mj}\alpha^{\ast}_{mj}}$,  then we   can write the squeezing parameter as follows :
\ben \label{sqp} \xi_{mj}(v)=\frac{2(-1)^{m}\chi \;t}{(1+j) \Delta} \frac{\widetilde{v}^{3}}{\omega''^{3}}(L_{j}^{(1)}(\frac{\widetilde{v}}{\omega''}))^2 e^{-\frac{\widetilde{v}}{\omega''}}, \een
with  $\widetilde{v}= v\Delta.$
The above equation represents the squeezing parameter which depends on the {\bf OAM} number $m$, the spectral order $j$, and the velocity of the two-modes $\mathrm{X}$ waves $v$.
Now we can show the normalized squeezing parameter modulus   $\vert\xi_{mj}^{(N)}\vert=\frac{\Delta\vert\xi_{mj}\vert}{4 \chi t}$ as a function of the normalized velocity $\frac{\widetilde{v}}{\omega''}$ for different values of $j$,($j= 0,1,2,3,4,5$), and fixing the time $t$ so that $\frac{4 \chi t }{\Delta}=1$, when $m=0$, as shown in Fig.1.

\begin{figure}[h]
\centering
\includegraphics[width=5cm,height=5cm]{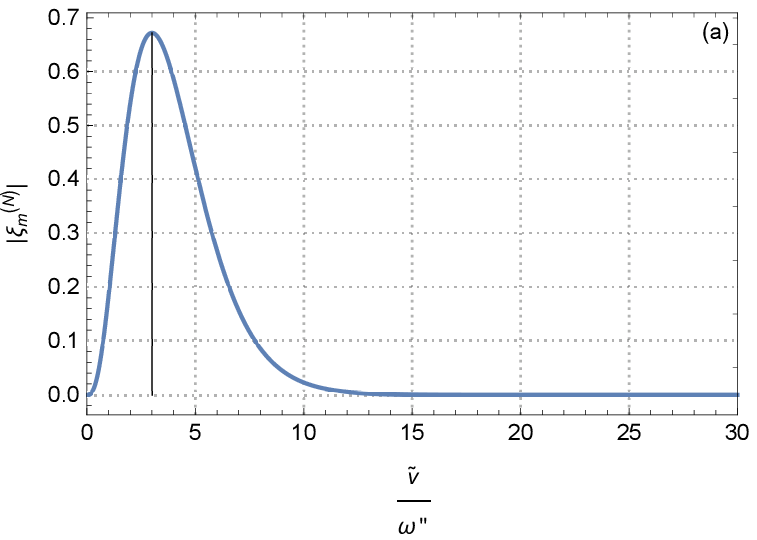}
\includegraphics[width=5cm,height=5cm]{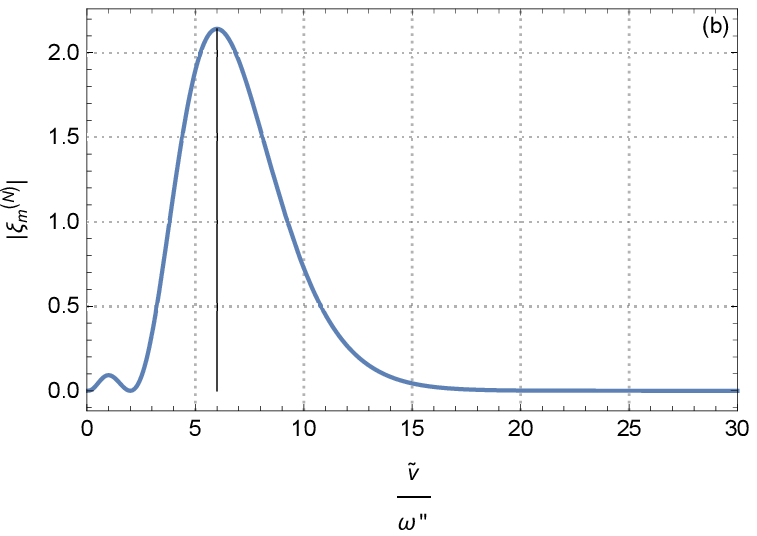}
\includegraphics[width=5cm,height=5cm]{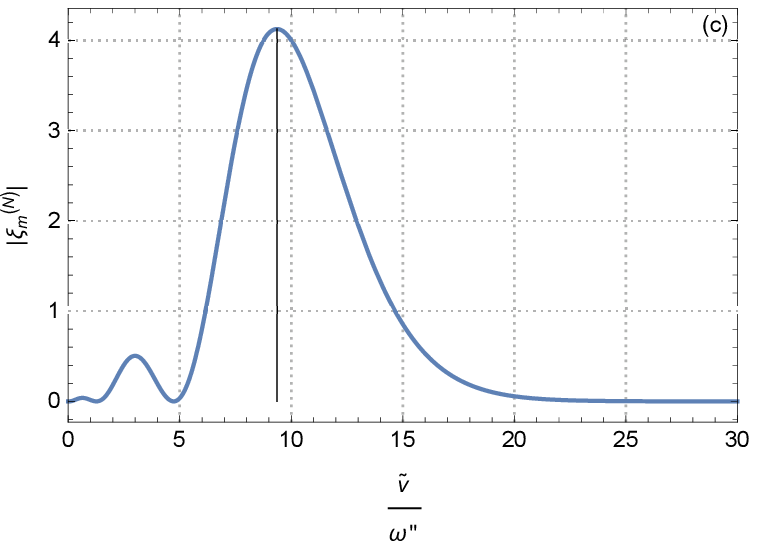}
\includegraphics[width=5cm,height=5cm]{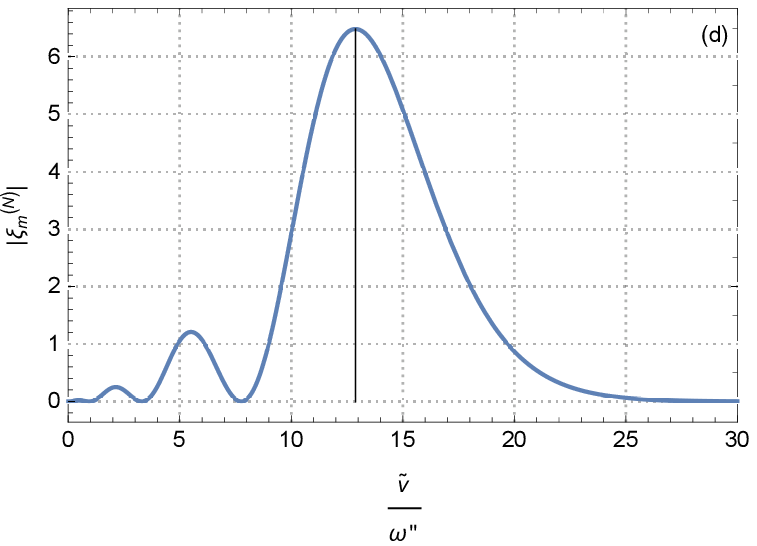}
\includegraphics[width=5cm,height=5cm]{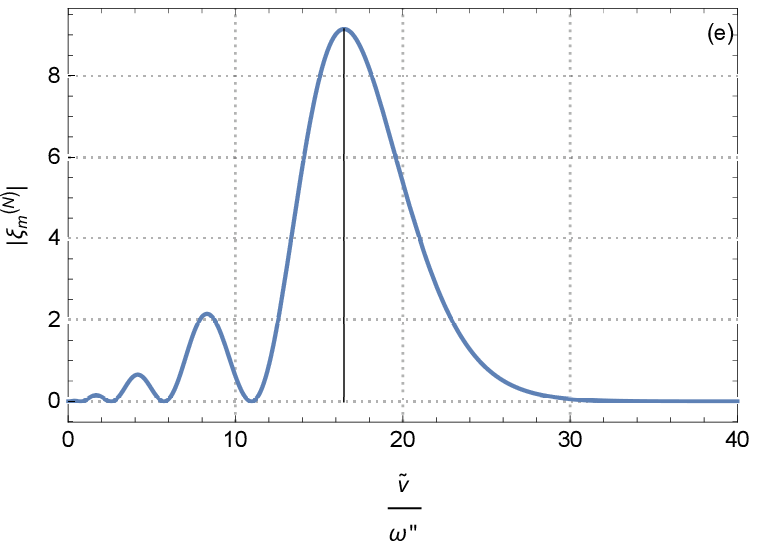}
\includegraphics[width=5cm,height=5cm]{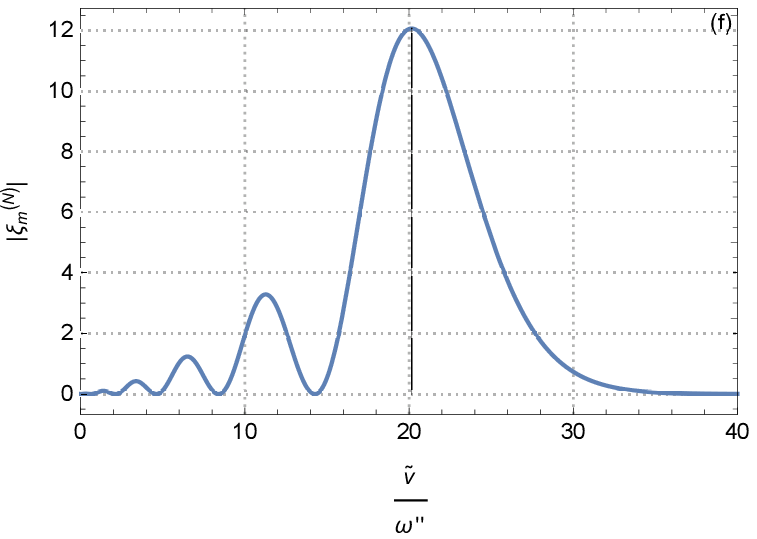}
\caption{ Plot of the normalized squeezing parameter modulus $|\xi_{mj}^{(N)}|=\frac{\Delta |\xi_{mj}|}{\chi}$ in function of the normalized velocity  $\frac{\widetilde{v}}{\omega''}$, for different values of the spectral order $j$ where,  (a)\; for $j=0$, \;(b) for $j=1$,\;(c) for $j=2$,\;(d) for $j=3$, \;(e) for $j=4$, \;(f) for $j=5,$ with fixing the time $t$ so that $\frac{4 \chi t}{\Delta}=1,$ and $m=0.$ }
\end{figure}
Fig.1 shows the dependence of the squeezing on the spectral order $j$ of $\mathrm{X}$ waves. This dependence can be explained in the following points:\begin{enumerate}
\item The maximum value of the squeezing parameter $\vert\xi_{mj}^{(N)}\vert$ and the corresponding normalized velocity $\frac{\widetilde{v}}{\omega''}$ increases as the spectral order $j$  increases.
\item The number of the peak values of squeezing depends on the spectral order $j$ and this number can be given by $j+1$, where the smallest peak is the first and the biggest one is the last.
\item The squeezing parameter depends on the velocity of $\mathrm{X}$ wave as shown in Eq.(\ref{sqp}) where the amount of squeezing parameter produced by {\bf SPDC} process will be maximized at an optimal velocity for every value of the spectral order $j$, and this velocity can be given by $v_{opt}=n_{j}\omega''/\Delta$, where $n_{0}=3$, $n_{1}=6$, $n_{2}=9.36$, $n_{3}=12.88$, $n_{4}=16.49$, $n_{5}=20.156$ for $j=0,1,2,3,4,5$ respectively as shown in Fig.1. Fig.1(a) shows that for $j=0$ as in ref.\cite{orn17}, and the corresponding optimal axicon angles for different values of $j$ can be given by $\cos\theta^{opt}_{j}=\Delta/n_{j}\lambda.$   For example, given a nondiffracting pulse with a duration of  $\Delta t= 8$\;fs, a carrier wave length of  $\lambda= 850$ nm  and  assuming that $\chi^{2}\simeq10^{-12}$ m/ V, the optimal axicon angles $\theta^{opt}_{j}$ that maximize the squeezing   for $j=0,1,2,3,4,5$ will take the following values respectively, $\theta^{opt}_{0}\simeq20^{\circ}$, $\theta^{opt}_{1}\simeq62^{\circ}$, $\theta^{opt}_{2}\simeq72^{\circ}$, $\theta^{opt}_{3}\simeq77^{\circ}$, $\theta^{opt}_{4}\simeq80^{\circ}$, $\theta^{opt}_{5}\simeq82^{\circ}.$
Then, we can evaluate the corresponded maximal squeezing parameters to get the following values respectively,  ($\xi^{(0)}_{m}\simeq100s^{-1}$, $\xi^{(1)}_{m}\simeq280s^{-1}$,  $\xi^{(2)}_{m}\simeq520s^{-1}$, $\xi^{(3)}_{m}\simeq820s^{-1}$, $\xi^{(4)}_{m}\simeq1160s^{-1}$, $\xi^{(5)}_{m}\simeq1550s^{-1}$).\end{enumerate}
	
To illustrate the effect of the  {\bf OAM} on the squeezing more, we can introduce the quadrature operators $\hat{X}_{mj}(v,t)=\hat{a}_{mj}(v,t)+\hat{a}_{mj}^{\dagger}(v,t),$ and $\hat{Y}_{mj}(v,t)=i [\hat{a}_{mj}^{\dagger}(v,t) - \hat{a}_{mj}(v,t)].$ For $e^{i\phi}= \alpha_{mj}/\beta_{mj} = 1$ or $\phi =0,$ we get $\hat{X}_{mj}(v,t)= e^{\xi_{mj}(v)} \hat{X}_{mj}(v,0),$ and $\hat{Y}_{mj}(v,t)= e^{-\xi_{mj}(v)} \hat{Y}_{mj}(v,0),$ and the variance of them is given by $\Delta \hat{X}_{mj}(v,t) = e^{\xi_{mj}(v)} \hat{Y}_{mj}(v,0).$ This shows that the {\bf SPDC} interaction Hamiltonian for the generalized $\mathrm{X}$ waves acts as  a two-modes squeeze operator which will be illustrated more in the next section. Noticeably, the effect of the {\bf OAM} number $m$ changes only the sign of the  squeezing parameter $\xi_{mj}(v)$ i.e. the squeezed quadrature changes depending on the parity of the orbital angular momentum number $m$. In particular, if $m$ is even number, $\xi_{mj}(v) > 0$, the squeezing accrues on the $Y$ quadrature. On the other hand, if $m$ is an odd number, $\xi_{mj}(v)< 0$ and the squeezing occurs on $X$ quadrature as in  \cite{orn17} for $j=0$, and for $j=1,2,3.$ this effect can be shown as in  Fig. 2.

\begin{figure}[h]
\centering
\includegraphics[width=7cm,height=4cm]{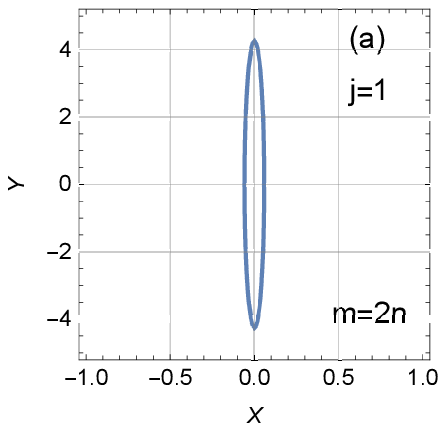}
\includegraphics[width=7cm,height=4cm]{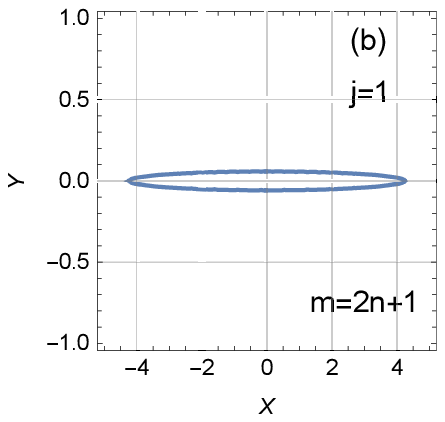}
\includegraphics[width=7cm,height=4cm]{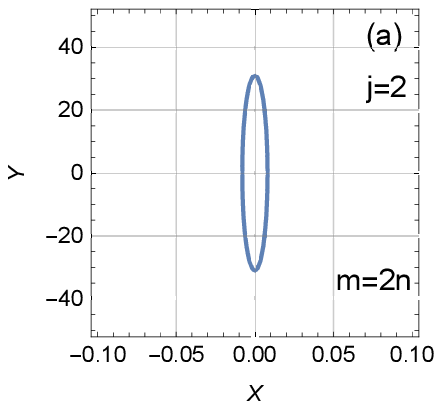}
\includegraphics[width=7cm,height=4cm]{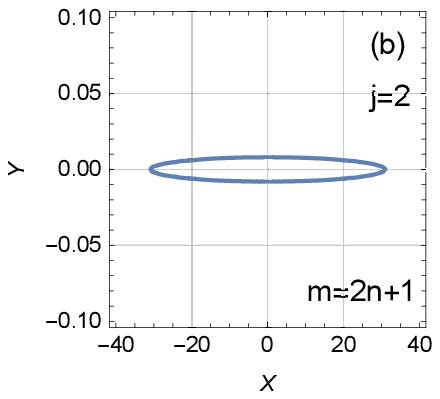}
\includegraphics[width=7cm,height=4cm]{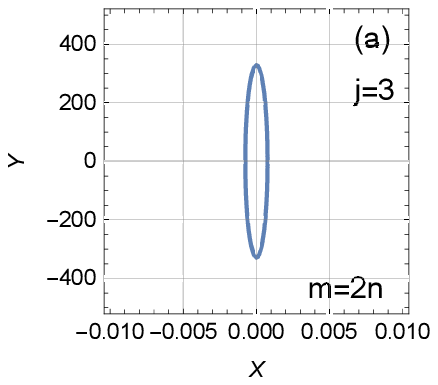}
\includegraphics[width=7cm,height=4cm]{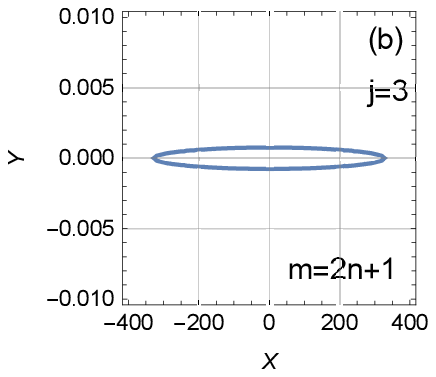}

\caption{Quadrature space representation of the squeezed down-converted state in the case of even [panel (a)] and odd [panel (b)] values of the {\bf OAM} number $m$ with a normalized velocity $\frac{\widetilde{v}}{\omega''}=6, 9.36, 12.88.$ for a spectral order $j=1,2,3,$ respectively.}
\end{figure}

\section{The two-modes squeezed state of $\mathrm{X}$ wave  and its entanglement}
In analogy to traditional case \cite{lon,ger}, we look for the eigenvalues of Hamiltonian operator, represented by Eq.(\ref{aa}). Let us introduce the $N$-particle states $|m,j,v;N\rangle$  as states with $N$ field excitation with velocity $v$  in the traveling mode (i.e. the  $\mathrm{X}$ wave) $\psi_{m,p}^{(v)}({\bf r},t)$ with energy $M v^2/2$ \cite{orn18}. The eigenstates $|m,p,v,N\rangle$ of the photon (particle) number operator $\hat{N}(v)= \hat{a}^{\dagger}_{mp}(v)\; \hat{a}_{mp}(v)$ are called Fock states, where $\hat{a}_{mp}(v)$ is the annihilation operator of the single mode in Eq.(\ref{anh}). These states are orthogonal \begin{equation}
	\langle n,q,u,N|m,p,v,N\rangle \equiv \delta_{nm}\; \delta_{pq} \;\delta(u-v),
	\end{equation} and complete. The number state $|m,p,v,N\rangle$ can be generated from the vacuum state (i.e. The eigenstate of $\hat{N}(v)$ with the eigenvalue equal to zero) as $$|m,p,v,N\rangle = \frac{[\hat{a}^{\dagger}_{mp}(v)]^{N}}{\sqrt{N!}} |0\rangle.$$   The expectation value of the electric field operator is zero on the particle number operator eigenstates, $$\langle m,p,v,N| \hat{A}(r,t)|m,p,v,N\rangle =0.$$ However, the Fock states $|m,p,v,N\rangle$ are not normalizable \cite{orn18}, since their representation in configuration state $\langle \vec{r},t|m,p,v,N\rangle = \psi^{(v)}_{mp}(r,t)$ carries infinite energy. The particle number states of the two modes $(A,B)$ with particle numbers $(N_{A},N_{B})$ are defined as   \begin{equation}|m,j,v;N_{A}\rangle_{A}=\frac{ [\hat{a}^{\dagger}_{mj}(v)]^{N_{A}}|0\rangle_{A}}{\sqrt{N_{A}!}}, \end{equation} and  \begin{equation}|-m,j,v,N_{B}\rangle_{B}=\frac{ [\hat{b}^{\dagger}_{-mj}(v)]^{N_{B}}|0\rangle_{B}}{\sqrt{N_{B}!}} \end{equation}
Together with the conditions for the vacuums, $\hat{a}_{mj}(v)|0\rangle_{A}=0$ and  $\hat{b}_{-mj}(v)|0\rangle_{B}=0$. \\ The operators $\hat{a}_{mj}(v)$ and $\hat{b}_{-mj}(v)$ obey the bosonic commutation relations in  Eq.(\ref{anh}). Now we can define the two-mode squeeze operator, which is unitary operator, using the normalized squeezing parameter $\xi_{mj}^{(N)}(v)= \frac{(-1)^{m}}{2(1+j)} \; \frac{\tilde{v}^{3}}{\omega''^{3}}\; (L^{(1)}_{j}(\frac{\tilde{v}}{\omega''}))^2\; e^{-\frac{\tilde{v}}{\omega''}},$ as \begin{equation} V_{\xi^{(N)}_{mj}}= \exp(e^{-i\phi} \; \xi_{mj}^{(N)} \; \hat{a}_{mj}(v) \; \hat{b}_{-mj}(v) - e^{i\phi} \; \xi_{mj}^{(N)}\;\hat{a}^{\dagger}_{mj}(v)\;\hat{b}^{\dagger}_{-mj}(v)). \end{equation} The two-mode squeezed vacuum {\bf TMSV} state can be obtained from the vacuum state as \begin{equation}\label{TMSV}|0,0,\xi^{(N)}_{mj}\rangle_{AB}=V_{\xi^{(N)}_{mj}}|0,0\rangle=[\cosh(\xi^{(N)}_{mj})]^{-1}\sum_{N=0}^{\infty}(e^{i\phi}\tanh(\xi^{(N)}_{mj}))^{N}|m,j,v,N;-m,j,v,N\rangle.\end{equation}  Also, we can define the annihilation operators crossing over the separation of part $A$ and part $B$ as \cite{Nam12},
$$\hat{A}_{\xi^{(N)}_{mj}}= V_{\xi^{(N)}_{mj}}\hat{a}_{mj}(v)V^{\dagger}_{\xi^{(N)}_{mj}}= \cosh(\xi^{(N)}_{mj})[\hat{a}_{mj}(v)+e^{i\phi}\tanh(\xi^{(N)}_{mj})\hat{b}^{\dagger}_{-mj}(v) ]$$
\begin{equation}\label{nanh} \hat{B}_{\xi^{(N)}_{mj}}= V_{\xi^{(N)}_{mj}}\hat{b}_{-mj}(v)V^{\dagger}_{\xi^{(N)}_{mj}}= \cosh(\xi^{(N)}_{mj})[\hat{b}_{-mj}(v)+e^{i\phi}\tanh(\xi^{(N)}_{mj})\hat{a}^{\dagger}_{mj}(v)].\end{equation}
The new field operators satisfy the bosonic commutation relations in Eq.(\ref{anh}) \begin{equation}[\hat{A}_{\xi^{(N)}_{mj}},\hat{A}^{\dagger}_{\xi^{(N)}_{m'j'}}]=[\hat{B}_{\xi^{(N)}_{mj}},\hat{B}^{\dagger}_{\xi^{(N)}_{m'j'}}]=\delta_{m,m'}\delta_{j,j'}\delta(v-v'),\end{equation}
 and
\begin{equation}[\hat{A}_{\xi^{(N)}_{mj}},\hat{B}_{\xi^{(N)}_{m'j'}}]=[\hat{A}_{\xi^{(N)}_{mj}},\hat{B}^{\dagger}_{\xi^{(N)}_{m'j'}}]=0.\end{equation}  The {\bf TMSV} state corresponds to the vacuum of the new field operators, as we have used $\hat{a}_{mj}(v)|0\rangle =0$, $\hat{b}_{-mj}(v)|0\rangle =0$, Eq.(\ref{TMSV}), Eq.(\ref{nanh}), and the unitary property of the squeeze operator, we get $\hat{A}_{\xi^{(N)}_{mj}}|0,0,\xi^{(N)}_{mj}\rangle=0$, and $\hat{B}_{\xi^{(N)}_{mj}}|0,0,\xi^{(N)}_{mj}\rangle=0$. \\ The two-mode number state is defined by the number state of nonlocal modes as \cite{Nam12},   \begin{equation} |N_A,N_B,\xi^{(N)}_{mj}\rangle = \frac{[\hat{A}^{\dagger}_{\xi^{(N)}_{mj}}]^{N_A}[\hat{B}^{\dagger}_{\xi^{(N)}_{mj}}]^{N_B}}{\sqrt{N_A!}\sqrt{N_B!}}|0,0,\xi_{\xi^{(N)}_{mj}}\rangle.\end{equation} In our case, we have $N_A=N_B=1$, so we obtain  \begin{equation}\label{squ1}|1,1,\xi^{(N)}_{mj}\rangle =\hat{A}^{\dagger}_{\xi^{(N)}_{mj}}\hat{B}^{\dagger}_{\xi^{(N)}_{mj}}|0,0,\xi^{(N)}_{mj}\rangle. \end{equation}  The state in the above equation represents a two-particles squeezed state of the two modes $\mathrm{X}$ wave.
  The superposition of the two particles corresponding to the two modes $\omega_{1}$ and $\omega_{2}$ see Eq.(\ref{stt1}) (when $u=v$) can be written as
\begin{equation} \label{stt2}|\psi^{(1)}(t) \rangle  =\sum_{mj} \int dv \; \mathcal{G}_{mj}(v)\; \mathcal{F}(v,t) \; \hat{a}^{\dagger}_{mj}(v)  \hat{b}^{\dagger}_{-mj}(v)|0,0\rangle, \end{equation} where $\mathcal{F}(v,t)=[ e^{i F(u)t}-1]$, and $\mathcal{G}_{mj}(v)= -i\; \omega(v) \chi_{mj}(2v)/F(v)$, with $\chi_{mj}(2v)$ as in Eq.(\ref{int2}). The superposition of the two particles squeezed state corresponds to the state in Eq.(\ref{stt2}) is
\begin{equation}|\psi^{(1)}_{\xi^{(N)}_{mj}}(t)\rangle =V_{\xi^{(N)}_{mj}} |\psi^{(1)}(t) \rangle  =\sum_{mj} \int dv \; \mathcal{G}_{mj}(v)\; \mathcal{F}(v,t)  V_{\xi^{(N)}_{mj}} \hat{a}^{\dagger}_{mj}(v)  \hat{b}^{\dagger}_{-mj}(v)|0,0\rangle. \end{equation}   After some calculations the state can be written in the form \begin{equation}\label{squ}|\psi^{(1)}_{\xi^{(N)}_{mj}}(t)\rangle =\sum_{mj} \int dv \; \mathcal{G}_{mj}(v)\; \mathcal{F}(v,t)  \hat{A}^{\dagger}_{\xi^{(N)}_{mj}} \hat{B}^{\dagger}_{\xi^{(N)}_{mj}}|0,0,\xi^{(N)}_{mj}\rangle, \end{equation}\\ which is a superposition of the squeezed states $|1,1,\xi^{(N)}_{mj}\rangle$ in Eq.(\ref{squ1}).
 Now, for calculating the entanglement of the state in Eq.(\ref{squ}), we consider the separable conditions based on the operators that form $SU(2)$ algebra and $SU(1,1)$ algebra \cite{Nam12}. The measurements of quadrature moments of fourth order are sufficient to verify the entanglement of this state. Let us define the operators,
$$J_{z}= \frac{1}{2} (N_{A}-N_{B}),$$ $$K_{x} = \frac{1}{2} (\hat{a}^{\dagger}_{mj}(v)\; \hat{b}^{\dagger}_{-mj}(v) + \hat{a}_{mj}(v)\; \hat{b}_{-mj}(v)),$$ \begin{equation} \label{opr}K_{y}= \frac{1}{2i}(\hat{a}^{\dagger}_{mj}(v) \hat{b}^{\dagger}_{-mj}(v) - \hat{a}_{mj}(v) \hat{b}_{-mj}(v))\end{equation}\\ Where $\hat{N}_{A}= \hat{a}^{\dagger}_{mj}(v)\hat{a}_{mj}(v)$ and $\hat{N}_{B}= \hat{b}^{\dagger}_{-mj}(v)\hat{b}_{-mj}(v)$ are number operators of the local modes.\\ The relations of the annihilation and creation operators $\hat{a}_{mj}(v)$,\; $\hat{a}^{\dagger}_{mj}(v)$, \;$\hat{b}_{-mj}(v)$, and $\hat{b}^{\dagger}_{-mj}(v)$ with the new field operators in Eq.(\ref{nanh}) can be represented by
$$\hat{a}_{mj}(v)= \cosh(\xi^{(N)}_{mj})(\hat{A}_{\xi^{(N)}_{mj}}+ e^{i\phi} \tanh(\xi^{(N)}_{mj}) \hat{B}^{\dagger}_{\xi^{(N)}_{mj}})$$
$$\hat{a}^{\dagger}_{mj}(v)= \cosh(\xi^{(N)}_{mj})(\hat{A}^{\dagger}_{\xi^{(N)}_{mj}}+ e^{-i\phi} \tanh(\xi^{(N)}_{mj}) \hat{B}_{\xi^{(N)}_{mj}})$$
$$\hat{b}_{-mj}(v)= \cosh(\xi^{(N)}_{mj})(\hat{B}_{\xi^{(N)}_{mj}}+ e^{i\phi} \tanh(\xi^{(N)}_{mj}) \hat{A}^{\dagger}_{\xi^{(N)}_{mj}})$$
\begin{equation} \label{anh2}\hat{b}^{\dagger}_{-mj}(v)= \cosh(\xi^{(N)}_{mj})(\hat{B}^{\dagger}_{\xi^{(N)}_{mj}}+ e^{-i\phi} \tanh(\xi^{(N)}_{mj}) \hat{A}_{\xi^{(N)}_{mj}})\end{equation}
The separable condition is \cite{Nam12}, \begin{equation}\label{cri}[\langle \Delta^{2} K_{y}\rangle-\frac{1}{4}]\langle \Delta^{2}J_{z}\rangle\geq \frac{1}{4}|\langle K_{x}\rangle|^{2} \end{equation}   Where $\langle \hat{O}\rangle=tr( \hat{O}\rho)$ denotes the expectation value of the observable $\hat{O}$ with respect to  state $\rho$, $\Delta\hat{O} =\hat{O}-\langle \hat{O}\rangle$, and $\langle\Delta^{2}\hat{O}\rangle=\langle\hat{O}^{2}\rangle-\langle \hat{O}\rangle^{2}$. The state we use in Eq.(\ref{cri}) is the superposition of simultaneous eigenstates of the number operators $\hat{A}^{\dagger}_{\xi^{(N)}_{mj}} \hat{A}_{\xi^{(N)}_{mj}}$ and $\hat{B}^{\dagger}_{\xi^{(N)}_{mj}}\hat{B}_{\xi^{(N)}_{mj}}$. From the definition of $J_{z}$ in Eq.(\ref{opr}), we have  \begin{equation} \hat{A}^{\dagger}_{\xi^{(N)}_{mj}} \hat{A}_{\xi^{(N)}_{mj}}-\hat{B}^{\dagger}_{\xi^{(N)}_{mj}} \hat{B}_{\xi^{(N)}_{mj}} = N_{A}- N_{B}= 2 J_{z} \end{equation} This implies that the {\bf TMS} number states are eigenstates of  $J_{z}$ and $\langle \Delta^{2}J_{z}\rangle=0.$\\  Also we can write
\begin{equation} K_{x}= \frac{1}{2}\cosh^{2}(\xi^{(N)}_{mj})[(\hat{A}^{\dagger}_{\xi^{(N)}_{mj}}+ \eta^{\ast}\hat{B}_{\xi^{(N)}_{mj}})(\hat{B}^{\dagger}_{\xi^{(N)}_{mj}}+\eta^{\ast}\hat{A}_{\xi^{(N)}_{mj}})+ (\hat{A}_{\xi^{(N)}_{mj}} + \eta \hat{B}^{\dagger}_{\xi^{(N)}_{mj}}) ( \hat{B}_{\xi^{(N)}_{mj}}+\eta\hat{A}^{\dagger}_{\xi^{(N)}_{mj}})],\end{equation}
with $\eta=e^{i\phi} \tanh(\xi^{(N)}_{mj})$.\\ Then the expectation value of $K_{x}$ for the {\bf TMS} number state of Eq.(\ref{squ}) can be given by
\begin{equation} \langle K_{x}\rangle = \sum_{mj}\int\frac{dv}{2}\cosh^{2}(\xi^{(N)}_{mj}) |\mathcal{G}_{mj}(v)|^2\; |\mathcal{F}(v,t)|^2  (\eta^{\ast}+\eta) \langle \hat{A}^{\dagger}_{\xi^{(N)}_{mj}}\hat{A}_{\xi^{(N)}_{mj}}+\hat{B^{\dagger}}_{\xi^{(N)}_{mj}}\hat{B}_{\xi^{(N)}_{mj}}+1\rangle\end{equation}  with $\langle \hat{A}^{\dagger}_{\xi^{(N)}_{mj}}\hat{A}_{\xi^{(N)}_{mj}}+\hat{B^{\dagger}}_{\xi^{(N)}_{mj}}\hat{B}_{\xi^{(N)}_{mj}}+1\rangle=3.$
From Eq.(\ref{ff}) we get the maximum value $|\mathcal{F}(v,t)|^2 =4$, thus \\\begin{equation}\langle K_{x}\rangle = 6 \sum_{mj}\int dv\cosh^{2}(\xi^{(N)}_{mj}) |\mathcal{G}_{mj}(v)|^2 (\eta^{\ast}+\eta)\end{equation}
where, $(\eta^{\ast}+\eta)=2 \cos(\phi) \tanh(\xi^{(N)}_{mj})$, which implies that $\langle K_{x}\rangle =0$, if $\phi=\frac{\pi}{2}$, this problem can be avoided by applying a local phase shift \cite{Nam12}, we obtain
$$\langle K_{x}\rangle = 12 \cos(\phi)\sum_{mj}\int dv  |\mathcal{G}_{mj}(v)|^2 \cosh(\xi^{(N)}_{mj})\sinh(\xi^{(N)}_{mj})$$
\begin{equation}= 6 \cos(\phi)\sum_{mj}\int dv  |\mathcal{G}_{mj}(v)|^2 \sinh(2 \xi^{(N)}_{mj}). \end{equation}  If we take $\phi=2n\pi, \; n=0,1,2,\cdots$, then  $\langle K_{x}\rangle$ will be maximized. This implies that $|\langle K_{x}\rangle|>0$ .Therefore, for the state in equation (\ref{squ}), the L.H.S. of Eq.(\ref{cri}) is equal to zero and its R.H.S. is positive. Hence, we can confirm the inseparability of the state in Eq.(\ref{squ}) from the violation of the condition in Eq.(\ref{cri}), means that we can take it as an indicator of quantifying the amount of the entanglement of squeezed states which can be detected by increasing $|\langle K_{x}\rangle|$. In other words we have found that  $|\langle K_{x}\rangle|$ increases as the spectral order $j$ increases, which indicats that the amount of entanglement of the state in Eq.(\ref{squ})  increases.

\section{Summary and comments}  To summarize, we obtain the velocity phase  matching, which can be used for determining the length of nonlinear crystal in the experiment setup to produce $\mathrm{X}$ wave with velocity $v$. We have introduced the relation between the squeezing parameter and the spectral order of $\mathrm{X}$ waves which shows that the maximal squeezing increases as the spectral order increases, and there exists an optimal velocity (i.e. axicon angle) that maximizes the amount of squeezing generated for each value of the spectral order. Moreover, we have constructed the squeeze operator of the two modes $\mathrm{X}$ wave, then we act by it on the two-particles state generated by {\bf SPDC} process to obtain the squeezed form of it. Finally, we have used the criterion of the quadrature moment of fourth order to verify the entanglement of the two modes squeezed state of $\mathrm{X}$ wave. We have observed  that the entanglement of the state increases as the spectral order increases.

\end{document}